\documentclass[pra,twocolumn,amssymb]{revtex4}
\usepackage{graphicx}
\usepackage{amsmath}

\begin{document}

\title{Geometric contribution to the Goldstone mode in spin-orbit coupled Fermi superfluids}

\author{M. Iskin}
\affiliation{Department of Physics, Ko\c{c} University, Rumelifeneri Yolu, 
34450 Sar\i yer, Istanbul, Turkey}

\date{\today}

\begin{abstract}

The so-called quantum metric tensor is a band-structure invariant whose measure 
corresponds to the quantum distance between nearby states in the Hilbert space, 
characterizing the geometry of the underlying quantum states.
In the context of spin-orbit coupled Fermi gases, we recently proposed that the 
quantum metric has a partial control over all those superfluid properties that depend 
explicitly on the mass of the superfluid carriers, i.e., the effective-mass tensor of 
the corresponding (two- or many-body) bound state. Here we scrutinize 
this finding by analyzing the collective phase and amplitude excitations at zero 
temperature. In particular to the Goldstone mode, we present extensive numerical 
calculations for the Weyl and Rashba spin-orbit couplings, revealing that, despite 
being small, the geometric contribution is solely responsible for the nonmonotonic 
evolution of the sound velocity in the BCS-BEC crossover.

\end{abstract}

\maketitle

\section{Introduction}
\label{sec:intro}

The quantum geometric tensor~\cite{provost80, berry89, thouless98} lies 
at the heart of modern solid-state and condensed-matter physics: while 
it describes the local geometry of the underlying Bloch states, it can be 
related to the global properties of some systems in most peculiar ways. 
For instance, its imaginary part corresponds to the Berry curvature 
measuring the emergent gauge field in momentum space, and it can be
associated with the topological invariants that are used to classify quantum 
Hall states or topological insulators~\cite{xiao10, qi11}. 
In contrast, its real part corresponds to the quantum metric tensor measuring 
the quantum distance/fidelity between nearby Bloch states, and it can be associated
with, e.g., the density and the effective-mass tensor of the superfluid (SF) 
carriers in spin-orbit coupled Fermi gases~\cite{iskin18a, iskin18b}, 
as well as with many other systems~\cite{resta11, tan19, yu18}. 

In the context of multi-band superconductivity, the new surge of theoretical
interest in quantum-metric effects dates back only to 2015, soon after its 
direct association with the geometric origin of superfluidity in topologically 
nontrivial flat bands~\cite{peotta15}. It was found that the SF weight (also
known as the SF stiffness tensor) has two distinct contributions based 
on the physical mechanisms involved. In addition to the conventional 
contribution that is controlled by the electronic spectra of the Bloch bands, 
there is also the so-called geometric contribution coming from the interband
processes~\cite{peotta15, liang17}. It later turns out that the quantum 
geometry has a partial control not only over the SF weight but also over all 
those SF properties that depend explicitly on the effective-mass tensor 
of the corresponding (two- or many-body) bound state~\cite{iskin18b}. 
There have been many studies on the subject exploring a variety of 
multi-band Hamiltonians, including most recently that of the twisted bilayer 
graphene~\cite{hu19, julku19, xie19}.

Motivated by the ongoing experimental progress~\cite{wang12, cheuk12, 
williams13, huang16, meng16} and given our recent 
findings~\cite{iskin18a, iskin18b,earlierworks}, 
here we consider a two-component Fermi gas with an arbitrary spin-orbit 
coupling (SOC), and analyze its collective excitations in the BCS-BEC crossover. 
We show that the velocity of the Goldstone mode has two contributions as well: 
in addition to the conventional intraband one that is controlled by the helicity 
spectrum, the geometric interband contribution is controlled by the quantum 
metric. We present extensive numerical calculations for the Weyl and Rashba 
SOCs revealing that, despite being much smaller than the conventional one, 
the geometric contribution is solely responsible for the nonmonotonic evolution 
of the sound velocity in the BCS-BEC crossover. 

The rest of this paper is organized as follows. In Sec.~\ref{sec:ce}, after introducing 
the many-body Hamiltonian, we first describe the effective-Gaussian action for the
phase- and amplitude-only degrees of freedoms, and then derive analytical 
expressions for the collective modes with a special emphasis on the distinction 
between the conventional and geometric contributions. Much of the theoretical 
details on the effective-action approach are deferred to the Appendix~\ref{sec:eaa}. 
The resultant Goldstone (phase) mode is analyzed Sec.~\ref{sec:numerics}, 
where we present numerical results for a 3D Fermi gas with a Weyl or a Rashba 
SOC, and for a 2D Fermi gas with a Rashba SOC for completeness. 
We end the paper with a brief summary of our conclusions in Sec.~\ref{sec:conc}.

\section{Collective Excitations}
\label{sec:ce}

Having a two-component ($\uparrow$ and $\downarrow$) Fermi gas with an 
arbitrary SOC that is driven across an $s$-wave Feshbach resonance in mind, 
we consider the Hamiltonian
\begin{align}
H &= \sum_\mathbf{k} (a_{\uparrow \mathbf{k}}^\dagger \, a_{\downarrow \mathbf{k}}^\dagger) 
\left[ \xi_{\mathbf{k}} \sigma_0  + \mathbf{d}_\mathbf{k} \cdot \boldsymbol{\sigma} \right]
\left( \begin{array}{c} a_{\uparrow \mathbf{k}} \\ a_{\downarrow \mathbf{k}} \end{array} \right) \nonumber \\
&- U \sum_{\mathbf{k} \mathbf{k'} \mathbf{q}} 
a_{\uparrow, \mathbf{k}+\mathbf{q}/2}^\dagger 
a_{\downarrow, -\mathbf{k}+\mathbf{q}/2}^\dagger
a_{\downarrow, -\mathbf{k'}+\mathbf{q}/2}
a_{\uparrow, \mathbf{k'}+\mathbf{q}/2},
\label{eqn:ham}
\end{align}
where $a_{\sigma \mathbf{k}}^\dagger$ ($a_{\sigma \mathbf{k}}$) creates (annihilates) 
a pseudospin-$\sigma$ fermion with momentum $\mathbf{k}$, i.e., in units of $\hbar \to 1$
the Planck constant. In the first line,
$
\xi_\mathbf{k} = \epsilon_\mathbf{k} - \mu
$ 
is the usual free-particle dispersion $\epsilon_\mathbf{k} = k^2/(2m)$ shifted by the 
chemical potential $\mu$,
$
\mathbf{d}_\mathbf{k} = \sum_i d_\mathbf{k}^i \boldsymbol{\widehat{i}}
$
is the SOC field with $\boldsymbol{\widehat{i}}$ denoting the unit vector along the 
$i = (x,y,z)$ direction, $\sigma_0$ is a $2 \times 2$ identity matrix, and 
$
\boldsymbol{\sigma} = \sum_i \sigma_i \boldsymbol{\widehat{i}}
$
is a vector of Pauli spin matrices. Our notation is such that
$
d_\mathbf{k}^i = \alpha_i k_i
$
corresponds to the Weyl SOC when $\alpha_i = \alpha$ for all $i$, and
to the Rashba SOC when $\alpha_z = 0$. We choose $\alpha \ge 0$ without 
the loss of generality. Thus, the one-body problem is described by the
Hamiltonian density
$
h_\mathbf{k}^0 = \epsilon_{\mathbf{k}} \sigma_0  + \mathbf{d}_\mathbf{k} \cdot \boldsymbol{\sigma},
$
determining the helicity spectrum
$
\epsilon_{s\mathbf{k}} = \epsilon_\mathbf{k} + s d_\mathbf{k}
$ 
and the helicity eigenstates $|s \mathbf{k} \rangle$ via the wave equation
$
h_\mathbf{k}^0 |s \mathbf{k} \rangle = \epsilon_{s\mathbf{k}} |s \mathbf{k} \rangle.
$
Here, $s = \pm$ labels the helicity bands, $d_\mathbf{k} = |\mathbf{d}_\mathbf{k}|$, 
and the unit vector
$
\widehat{\mathbf{d}}_\mathbf{k} = \mathbf{d}_\mathbf{k}/d_\mathbf{k}
$
plays an important role throughout the paper.

In the second line of Eq.~(\ref{eqn:ham}), $U \ge 0$ corresponds to the strength 
of the zero-ranged density-density interactions between $\uparrow$ and 
$\downarrow$ particles, and we tackle this term by making use of the 
imaginary-time functional path-integral formalism for the paired fermions. 
This formalism offers a systematic approach to go beyond the mean-field theory
as the fluctuation effects can be incorporated into the effective action order 
by order as a power-series expansion around the stationary 
saddle-point~\cite{carlos97, iskin05, he12, shenoy12}.
As shown in the Appendix, the zeroth-order (i.e., the saddle-point) action 
coincides with the mean-field one, and can be written as
\begin{align}
S_0 = \frac{\Delta_0^2}{T U} 
+ \sum_{s \mathbf{k}} \left\lbrace \frac{\xi_{s\mathbf{k}} - E_{s\mathbf{k}}}{2T} 
+ \ln [f(-E_{s\mathbf{k}})] \right\rbrace,
\label{eqn:S0}
\end{align}
where
$
\Delta_0 = U\langle a_{\uparrow \mathbf{k}} a_{\downarrow,-\mathbf{k}} \rangle
$
is the saddle-point order parameter with $\langle \dots \rangle$ denoting 
a thermal average, $T$ is the temperature in units of $k_B \to 1$ the 
Boltzmann constant,
$
E_{s\mathbf{k}} = \sqrt{\xi_{s\mathbf{k}}^2+\Delta_0^2}
$
is the quasiparticle energy spectrum associated with
$
\xi_{s\mathbf{k}} = \epsilon_{s\mathbf{k}} - \mu,
$
and $f(x) = 1/(e^{x/T} + 1)$ is the Fermi-Dirac distribution. Here, we take 
$\Delta_0$ to be a real parameter without the loss of generality, and 
determine its value by imposing the saddle-point condition 
$\partial S_0/\partial \Delta_0 = 0$ on the order parameter, leading to
$
1 = U \sum_{s \mathbf{k}} [1 - 2f(E_{s\mathbf{k}})]/(4E_{s \mathbf{k}}).
$
Furthermore, $\mu$ is determined by the thermodynamic relation
$
N_0 = - T \partial S_0 /\partial \mu
$
for the number of particles, leading to
$
N_0 = \sum_{s \mathbf{k}} \{1 - (\xi_{s\mathbf{k}} /E_{s\mathbf{k}}) [1 - 2f(E_{s\mathbf{k}})]\}/2.
$
These saddle-point equations constitute the self-consistent BCS-BEC 
crossover theory for pairing, and they coincide with those of the mean-field 
approximation. The coupled solutions for $\Delta_0$ and $\mu$ have served 
as a reliable guide for the cold-atom community, providing a qualitative 
description of the SF state at sufficiently low temperatures across an 
$s$-wave Feshbach resonance~\cite{carlos97, iskin05, he12, shenoy12}.

Going beyond $S_0$ in the Appendix, we first introduce the bosonic field
$
\Delta_q = \Delta_0 + \Lambda_q,
$
where $\Lambda_q$ corresponds to the fluctuations of the order parameter
around its stationary value $\Delta_0$ with the collective index 
$q = (\mathbf{q}, \mathrm{i}\nu_n)$ denoting both the pair momentum $\mathbf{q}$ 
and the bosonic Matsubara frequency $\nu_n = 2\pi n T$, and then 
derive the effective-Gaussian action $S_\mathrm{Gauss} = S_0 + S_2$. 
Note that the first order term $S_1$ vanishes thanks to the saddle-point 
condition on the order parameter. Furthermore, it is physically insightful to 
separate the phase and amplitude degrees of freedoms through the unitary 
transformation
$
\Lambda_q = (\lambda_q + \mathrm{i} \theta_q)/\sqrt{2},
$
leading to a phase-only action $S_2^\theta$ and to a amplitude-only one
$S_2^\lambda$ that are of the form
\begin{align}
\label{eqn:S2theta}
S_2^\theta &= \frac{1}{2T} \sum_q \theta_q^* 
\left[ M_{q,E}^{11} - M_q^{12} - \frac{(M_{q,O}^{11})^2}{M_{q,E}^{11}+M_q^{12}} \right] \theta_q, \\ 
S_2^\lambda &= \frac{1}{2T} \sum_q \lambda_q^* 
\left[ M_{q,E}^{11} + M_q^{12} - \frac{(M_{q,O}^{11})^2}{M_{q,E}^{11}-M_q^{12}} \right] \lambda_q
\label{eqn:S2lambda}.
\end{align}
Here, since $M_{q,O}^{11}$ couples the phase and amplitude modes, the phase-only 
action is derived by integrating out the amplitude fields, and vice versa~\cite{iskin05}. 
While the thermal expressions for the matrix elements $M_q^{ij}$ of the 
inverse-fluctuation propagator $\mathbf{M}_q$ are given in the Appendix, 
here we restrict ourselves to the collective excitations on top of the thermal 
ground state. Thus, setting $T = 0$, and denoting
$\xi_{s, \mathbf{k}+\mathbf{q}/2}$ by $\xi$;
$\xi_{s', -\mathbf{k}+\mathbf{q}/2}$ by $\xi'$;
$E_{s, \mathbf{k}+\mathbf{q}/2}$ by $E$;
$E_{s', -\mathbf{k}+\mathbf{q}/2}$ by $E'$; and
$\widehat{\mathbf{d}}_{\pm \mathbf{k}+\mathbf{q}/2}$ by $\widehat{\mathbf{d}}_\pm$,
a compact way to write the matrix elements is~\cite{iskin05}
\begin{align}
M_{q,E}^{11} &= \frac{1}{U} + \sum_{ss' \mathbf{k}} \frac{(\xi \xi' + E E')(E+E')
(1 - ss' \widehat{\mathbf{d}}_+ \cdot \widehat{\mathbf{d}}_-)}
{8E E'[(\mathrm{i}\nu_n)^2 - (E+E')^2]}, \\
M_{q,O}^{11} &= \sum_{ss' \mathbf{k}} \frac{(\xi E' + E \xi')\mathrm{i}\nu_n 
(1 - ss' \widehat{\mathbf{d}}_+ \cdot \widehat{\mathbf{d}}_-)}
{8E E'[(\mathrm{i}\nu_n)^2 - (E+E')^2]}, \\
M_q^{12} &= -\sum_{ss' \mathbf{k}} \frac{\Delta_0^2(E+E')
(1 - ss' \widehat{\mathbf{d}}_+ \cdot \widehat{\mathbf{d}}_-)}
{8E E'[(\mathrm{i}\nu_n)^2 - (E+E')^2]}.
\end{align}
Here, $M_{q,E}^{11}$ and $M_q^{12}$ are even functions of $\mathrm{i}\nu_n$, but 
$M_{q,O}^{11}$ is an odd one. 

The dispersions $\omega_\mathbf{q}$ for the collective modes are determined 
by the poles of the propagator matrix $\mathbf{M}_q^{-1}$ for the pair fluctuation 
fields, i.e., $\omega_\mathbf{q}^\theta$ for the phase mode and 
$\omega_\mathbf{q}^\lambda$ for the amplitude one are determined, respectively,
by setting $[ \dots ] = 0$ in Eqs.~(\ref{eqn:S2theta}) and~(\ref{eqn:S2lambda}) 
after an analytic continuation $\mathrm{i}\nu_n \to \omega + i0^+$ to the real 
axis~\cite{carlos97, iskin05, he12, shenoy12}. 
For this purpose, it is sufficient to retain terms that are up to quadratic orders 
in their small $\mathbf{q}$ and $\omega$ expansions~\cite{collnote}, i.e., 
$
M_{q,E}^{11}+M_q^{12} = A + \sum_{ij} C_{ij} q_i q_j - D\omega^2 + \cdots;
$
$
M_{q,E}^{11}-M_q^{12} = \sum_{ij} Q_{ij} q_i q_j - R\omega^2 + \cdots;
$
and
$
M_{q,O}^{11} = -B\omega + \cdots.
$
This calculation leads to a gapless phase (Goldstone) mode
$
(\omega_\mathbf{q}^\theta)^2 = \sum_{ij} x_{ij} q_i q_j
$
with 
$
x_{ij} = Q_{ij}/(R+ B^2/A)
$
for the phase-only action, and to a gapped amplitude (Higgs) mode
$
(\omega_\mathbf{q}^\lambda)^2 = \omega_0^2 + \sum_{ij} y_{ij} q_i q_j
$
with 
$
\omega_0 = \sqrt{(A + B^2/R)/D}
$
and
$
y_{ij} = C_{ij}/D + Q_{ij}/(R + R^2A/B^2)
$
for the amplitude-only action. Notice that the velocities of the modes have 
a tensor structure in general, and can be anisotropic depending on the 
$\mathbf{k}$-space structure of the SOC. We are interested only in the 
low-energy Goldstone mode at $T = 0$ since this mode is not damped 
for all $U \ne 0$ provided that the quasiparticle excitations $E_{s\mathbf{k}}$ 
are gapped for all $\Delta_0 \ne 0$.

It turns out that the general expressions for the nonkinetic-expansion 
coefficients can simply be written as a sum over their well-known 
counterparts for the usual one-band problem, i.e.,
$
A = \sum_{s \mathbf{k}} \Delta_0^2/(4E_{s\mathbf{k}}^3),
$
$
B = \sum_{s \mathbf{k}} \xi_{s\mathbf{k}}/(8E_{s\mathbf{k}}^3),
$
$
D = \sum_{s \mathbf{k}} \xi_{s\mathbf{k}}^2/(16E_{s\mathbf{k}}^5),
$
and
$
R = \sum_{s \mathbf{k}} 1/(16E_{s\mathbf{k}}^3)
$
are due entirely to intraband mechanisms~\cite{carlos97, iskin05, he12, shenoy12}.
Motivated by the recent findings~\cite{iskin18a, iskin18b, earlierworks}, and 
to gain more physical insight, we divide the kinetic-expansion coefficients into 
two distinct contributions
$
Q_{ij} = Q_{ij}^\mathrm{intra} + Q_{ij}^\mathrm{inter}
$
and 
$
C_{ij} = C_{ij}^\mathrm{intra} + C_{ij}^\mathrm{inter},
$
based on their physical origins, i.e., depending on whether intraband 
or interband processes are involved. For instance, a compact way to 
write their intraband (i.e., conventional) contributions is
\begin{align}
Q_{ij}^\mathrm{intra} &= \sum_{s \mathbf{k}} \frac{1}{16 E_{s\mathbf{k}}^3}
\frac{\partial \xi_{s\mathbf{k}}}{\partial k_i} \frac{\partial \xi_{s\mathbf{k}}}{\partial k_j}, \\
C_{ij}^\mathrm{intra} &= \sum_{s \mathbf{k}} \frac{1}{16 E_{s\mathbf{k}}^3} 
\left( 1 - \frac{5\Delta_0^2 \xi_{s\mathbf{k}}^2}{E_{s\mathbf{k}}^4} \right)
\frac{\partial \xi_{s\mathbf{k}}}{\partial k_i} \frac{\partial \xi_{s\mathbf{k}}}{\partial k_j}.
\end{align}
We note that these terms can be transformed into their more familiar but more 
involved forms~\cite{carlos97, sdMforms} through first writing
$
\partial [(\partial \xi_{s\mathbf{k}}/\partial k_j)/E_{s\mathbf{k}}^3]/\partial k_i 
= [\partial^2 \xi_{s\mathbf{k}}/(\partial k_i \partial k_j)]/E_{s\mathbf{k}}^3
- 3 (\partial \xi_{s\mathbf{k}}/\partial k_i) (\partial \xi_{s\mathbf{k}}/\partial k_j) 
\xi_{s\mathbf{k}}/E_{s\mathbf{k}}^5,
$
and then identifying 
$
\sum_\mathbf{k} \xi_{s\mathbf{k}} \partial [(\partial \xi_{s\mathbf{k}}/\partial k_j)/E_{s\mathbf{k}}^3]/\partial k_i 
\equiv - \sum_\mathbf{k}  (\partial \xi_{s\mathbf{k}}/\partial k_i) (\partial \xi_{s\mathbf{k}}/\partial k_j)/E_{s\mathbf{k}}^3
$
after an integration by parts, and hence the name conventional. On the other hand, 
a compact way to write their interband (i.e., geometric) contributions is
\begin{align}
Q_{ij}^\mathrm{inter} &= - \sum_{s \mathbf{k}} \frac{d_\mathbf{k}}{8 s\xi_\mathbf{k} E_{s\mathbf{k}}}
g_\mathbf{k}^{ij}, \\
C_{ij}^\mathrm{inter} &= - \sum_{s \mathbf{k}} \frac{d_\mathbf{k}}{8 s\xi_\mathbf{k} E_{s\mathbf{k}}}
\left( 1 + \frac{2\Delta_0^2}{d_\mathbf{k}^2} \right)
g_\mathbf{k}^{ij},
\end{align}
where 
$
g_{\mathbf{k}}^{ij} = (1/2)
\lim_{\mathbf{q} \to \mathbf{0}} 
\partial^2(\widehat{\mathbf{d}}_+ \cdot \widehat{\mathbf{d}}_-)
/ (\partial q_i \partial q_j)
$
or equivalently
$
g_{\mathbf{k}}^{ij} = (\partial \widehat{\mathbf{d}}_\mathbf{k}/\partial k_i) 
\cdot (\partial \widehat{\mathbf{d}}_\mathbf{k}/\partial k_j)/2
$
is the total quantum metric tensor of the helicity bands~\cite{iskin18a, iskin18b}. 
Alternatively, the quantum metric can be written as
$
g_\mathbf{k}^{ij} = [\sum_\ell (\partial d_\mathbf{k}^\ell/\partial k_i) 
(\partial d_\mathbf{k}^\ell/\partial k_j) 
- (\partial d_\mathbf{k}/\partial k_i) (\partial d_\mathbf{k}/\partial k_j)]/(2d_\mathbf{k}^2),
$
reducing to
$
g_\mathbf{k}^{ij} = \alpha_i \alpha_j (d_\mathbf{k}^2 \delta_{ij} - d_\mathbf{k}^i d_\mathbf{k}^j)
/(2d_\mathbf{k}^4)
$
for the specific case when $d_\mathbf{k}^i = \alpha_i k_i$. Since the interband
contributions are directly controlled by the geometry of the underlying quantum 
states in the parameter ($\mathbf{k}$) space, we refer to them as the geometric ones.

All of the coefficients $A$, $B$, $C_{ij}$, etc., are derived at $T = 0$ for arbitrary 
$U$ and $\mathbf{d}_\mathbf{k}$, and can be used both in two and three dimensions. 
Next we present their specific application to three systems that are of immediate 
experimental and/or theoretical interest:
(i) a 3D Fermi gas with a Weyl SOC, 
(ii) a 3D Fermi gas with a Rashba SOC, and 
(iii) a 2D Fermi gas with a Rashba SOC.

\section{Numerical Results}
\label{sec:numerics}

In this paper, our primary interest is in the cooperation and/or competition 
between the conventional intraband contribution
$
x_{ij}^\mathrm{intra} = Q_{ij}^\mathrm{intra}/(R+ B^2/A)
$
and the geometric interband one
$
x_{ij}^\mathrm{inter} = Q_{ij}^\mathrm{inter}/(R+ B^2/A)
$
on the velocity of the Goldstone mode. For this purpose, one first needs to 
obtain a self-consistent solution for $\Delta_0$ and $\mu$ by coupling the 
saddle-point condition and the saddle-point number equation, and then plug 
these solutions into the coefficients.

\begin{figure}[htbp]
\includegraphics[scale=0.45]{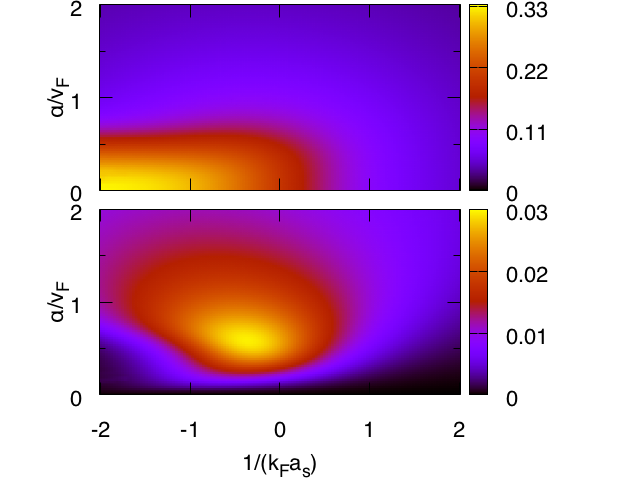}
\caption{(color online)
\label{fig:3DWeyl}
A 3D Fermi gas with Weyl SOC is mapped at $T = 0$ in the plane of 
two-body scattering length $a_s$ and SOC strength $\alpha$. 
Square of the sound velocity $x_0/v_F^2$ is shown (upper panel) along 
with its geometric contribution $x_0^\mathrm{inter}/v_F^2$ (lower panel).
}
\end{figure}

For instance, in Fig.~\ref{fig:3DWeyl}, we show $x_{ij} = x_0 \delta_{ij}$ for a 
3D Fermi gas with a Weyl SOC, where 
$
\mathbf{d}_\mathbf{k}  = \alpha \mathbf{k},
$ 
and therefore,
$
x_0 = x_0^\mathrm{intra} + x_0^\mathrm{inter}
$ 
is isotropic in the entire $\mathbf{q}$ space with $\delta_{ij}$ a Kronecker-delta. 
In accordance with the cold-atom literature, the two-body $s$-wave scattering 
length $a_s$ in vacuum is substituted in this figure for $U$ via the usual relation
$
V/U = -mV/(4\pi a_s) + \sum_\mathbf{k} 1/(2\epsilon_\mathbf{k}),
$
where $V$ is the volume of the system. Furthermore, we choose
$
N = k_F^3 V/(3\pi^2)
$
for the number of particles, and use the Fermi momentum $k_F$ and the 
Fermi velocity $v_F = k_F/m$ as the relevant length and velocity scales in our 
numerical calculations. This is such that increasing $U$ from $0$ changes the 
dimensionless parameter $1/(k_F a_s)$ continuously from $-\infty$ to 
$0$ to $+\infty$ in the BCS-BEC crossover.

In the weak-SOC limit when $\alpha/v_F \ll 1$, we recover the usual problem 
where $x_0/v_F^2$ is known to be a monotonically decreasing function 
of $1/(k_Fa_s)$ with the well-known limits, i.e., $1/3$ in the BCS regime 
and $k_F a_s/ (3\pi)$ in the BEC regime~\cite{carlos97, iskin05}. 
Note that since the geometric contribution $x_0^\mathrm{inter}$ is originating 
from the interband processes, it plays a negligible role when $\alpha/v_F \to 0$. 
On the other hand, for intermediate $\alpha/v_F$ values, we find that 
$x_0/v_F^2$ is a nonmotonic function of $1/(k_Fa_s)$, exhibiting a 
noticeable peak in the crossover region around the unitarity~\cite{he12, shenoy12}. 
This result is quite a surprise given that all of the ground-state properties 
are found to be monotonic functions of $1/(k_Fa_s)$ for the usual BCS-BEC 
crossover problem with $s$-wave interactions~\cite{carlos97}. In comparison 
to the dominant contribution from $x_0^\mathrm{intra}$, our study reveals that, 
despite being small, the contribution from $x_0^\mathrm{inter}$ is sizeable 
enough to cause such a significant difference in $x_0$. Thus, even though a 
direct observation of $x_0^\mathrm{inter}$ may not be possible, realization
of a nonmonotonic $x_0$ may be considered as the ultimate evidence 
for its subleading role.

\begin{figure}[htbp]
\includegraphics[scale=0.45]{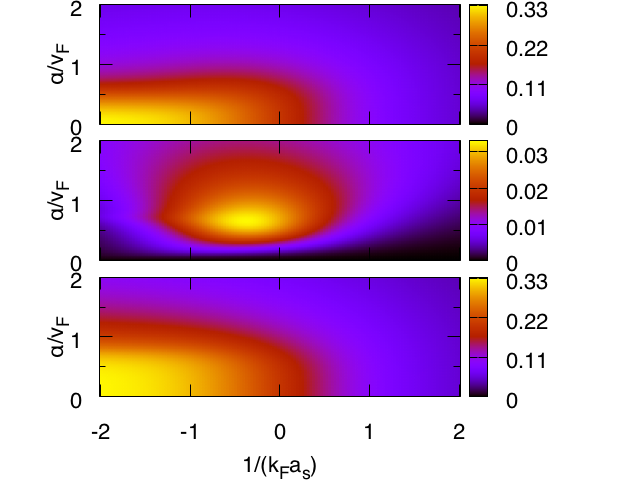}
\caption{(color online)
\label{fig:3DRashba}
A 3D Fermi gas with Rashba SOC is mapped at $T = 0$ in the plane of 
two-body scattering length $a_s$ and SOC strength $\alpha$. 
Square of the sound velocities $x_\perp/v_F^2$ and $x_{zz}/v_F^2$ 
are shown (upper and lower panels) along with the geometric contribution 
$x_\perp^\mathrm{inter}/v_F^2$ (middle panel) of the former.
}
\end{figure}

Similarly, in Fig.~\ref{fig:3DRashba}, we show $x_{ij} = x_{ii} \delta_{ij}$ 
for a 3D Fermi gas with a Rashba SOC, where 
$
\mathbf{d}_\mathbf{k}  = \alpha \mathbf{k}_\perp
$ 
with 
$
\mathbf{k}_\perp = k_x \boldsymbol{\widehat{x}} + k_y \boldsymbol{\widehat{y}},
$
and therefore,
$
x_\perp = x_\perp^\mathrm{intra} + x_\perp^\mathrm{inter}
$ 
is still isotropic in $q_xq_y$ plane but it differs from that $x_{zz} = x_{zz}^\mathrm{intra}$ 
of the $q_z$ direction with $x_{zz}^\mathrm{inter} = 0$. 
Note that since $g_\mathbf{k}^{iz} = 0$ for a Rashba SOC, all of the matrix 
elements $x_{iz}^\mathrm{inter}$ trivially vanish, leading to the aforementioned 
anisotropy. For intermediate $\alpha/v_F$ values, we again find that $x_\perp/v_F^2$ 
is a nonmonotonic function of $1/(k_Fa_s)$, exhibiting a noticeable peak in 
the crossover region around the unitarity. 
However, $x_{zz}/v_F^2$ turns out to be a monotonic one no matter what 
$\alpha/v_F$ is. Thus, the dramatic difference between $x_\perp$ and $x_{zz}$ 
may be used as an ultimate evidence for the presence of an interband contribution 
to the former, reflecting the geometry of the underlying quantum states.

\begin{figure}[htbp]
\includegraphics[scale=0.45]{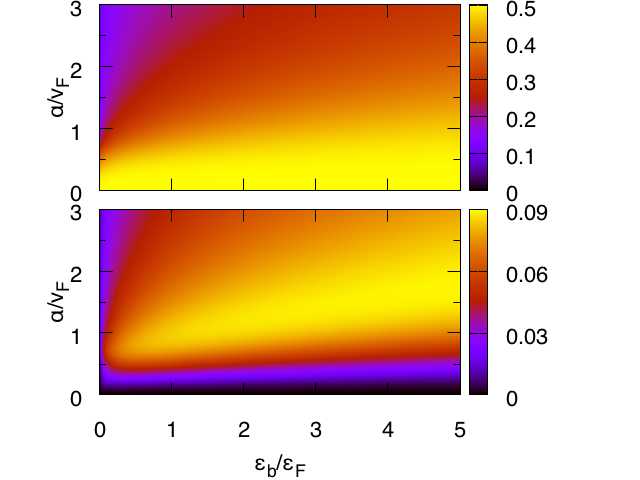}
\caption{(color online)
\label{fig:2DRashba}
A 2D Fermi gas with Rashba SOC is mapped at $T = 0$ in the plane of 
two-body binding energy $\epsilon_b$ and SOC strength $\alpha$. 
Square of the sound velocity $x_0/v_F^2$ is shown (upper panel) along 
with its geometric contribution $x_0^\mathrm{inter}/v_F^2$ (lower panel).
}
\end{figure}

For completeness, in Fig.~\ref{fig:2DRashba}, we show $x_{ij} = x_0 \delta_{ij}$ 
for a 2D Fermi gas with a Rashba SOC, where 
$
\mathbf{d}_\mathbf{k}  = \alpha \mathbf{k},
$ 
and therefore,
$
x_0 = x_0^\mathrm{intra} + x_0^\mathrm{inter}
$ 
is isotropic in the entire $\mathbf{q}$ space. In accordance with the cold-atom literature, 
the two-body binding energy $\varepsilon_b \ge 0$ in vacuum is substituted in this 
figure for $U$ via the usual relation
$
A/U = \sum_\mathbf{k} 1/(2\epsilon_\mathbf{k} + \varepsilon_b),
$
where $A$ is the area of the system. Furthermore, we choose
$
N = k_F^2 A/(2\pi),
$
and use the Fermi energy $\epsilon_F = k_F^2/(2m)$ as the relevant energy 
scale in our numerical calculations. This is such that increasing $U$ 
from $0$ increases $\varepsilon_b/\epsilon_F$ continuously from $0$ to $\infty$ 
in the BCS-BEC crossover.  
In the weak-SOC limit when $\alpha/v_F \ll 1$, we again recover the usual 
problem where $x_0/v_F^2$ is known to be $1/2$ for all $\varepsilon_b/\epsilon_F$, 
i.e., it remains a constant in the entire BCS-BEC crossover~\cite{iskin05}. 
In contrast to the 3D results shown in Figs.~\ref{fig:3DWeyl} and~\ref{fig:3DRashba} 
where the geometric contribution $x_0^\mathrm{inter}$ plays important roles 
only in the crossover region with a rapid decay in the BCS and BEC regimes, 
here $x_0^\mathrm{inter}$ retains its strength for most $\varepsilon_b/\epsilon_F$
values except for the extreme BCS limit when $\varepsilon_b/\epsilon_F \ll 1$. 
Even then we find that the fraction $x_0^\mathrm{inter}/x_0$ does not grow 
much, and saturates around a quarter.

\section{Conclusions}
\label{sec:conc}

In summary, here we considered a two-component Fermi gas with an arbitrary 
SOC, and analyzed its collective excitations in the BCS-BEC crossover. 
Motivated by the recent findings~\cite{iskin18a, iskin18b}, we divided the 
velocity of the Goldstone mode into two distinct contributions based on the
physical mechanisms involved. 
While the conventional (intraband) contribution is controlled by the 
$\mathbf{k}$-space derivatives of the helicity (one-body) spectrum, the 
geometric (interband) one is controlled by the $\mathbf{k}$-space derivatives 
of the helicity eigenstates or the so-called quantum metric tensor characterizing 
the geometry of the underlying quantum states. We presented extensive 
numerical calculations for the Weyl and Rashba SOCs revealing that, despite 
being much smaller than the conventional one, the geometric contribution is 
solely responsible for the nonmonotonic evolution of the mode velocity in 
the BCS-BEC crossover. 

These findings for the Goldstone mode in spin-orbit coupled Fermi gases are in 
complete agreement with our recent works that the quantum metric tensor has a 
subleading control over all those SF properties that depend explicitly on the 
effective-mass tensor of the corresponding (two- or many-body) bound 
state~\cite{iskin18a, iskin18b}. 
We note that whether the quantum-metric contribution can be isolated from the 
total is far from being clear with the current experimental techniques involving 
spin-orbit coupling. However, in the same spirit, given that the starting 
Hamiltonian Eq.~(\ref{eqn:ham}) is quite generic, our analytical expressions 
for the Goldstone mode may also find direct applications in other contexts 
that exhibit a two-band band structure. 
In particular, it is desirable to study the Goldstone mode in a narrow- or a 
flat-band system~\cite{iskin19a, iskin19b}, for which the geometric 
contribution is expected to take the lead. As a final remark, our work suggests 
that an analogous contribution to the collective excitations must be present 
in many other multi-band systems including the twisted bilayer graphene.

\begin{acknowledgments}
This work is supported by the funding from T{\"U}B{\.I}TAK Grant No. 1001-118F359.
\end{acknowledgments}

\appendix

\section{Effective-Action Approach}
\label{sec:eaa}

The imaginary-time functional path-integral formalism allows a systematic derivation 
of the effective action in the form of a power-series expansion of the fluctuations 
of the order parameter around its saddle-point 
value~\cite{carlos97, iskin05, he12, shenoy12}. In this approach, 
to decouple the interaction term that is quartic in the fermionic degrees 
of freedom, one first uses a Hubbard-Stratanovich transformation at the expense 
of introducing a bosonic field $\Delta_q$, and then integrates out the remaining 
terms that are quadratic in the fermionic degrees of freedom. Finally, by decomposing 
$\Delta_q = \Delta_0 + \Lambda_q$ in terms of a $q$-independent stationary 
field $\Delta_0$ and $q$-dependent fluctuations around it, one may in principle 
obtain an effective action at the desired order in $\Lambda_q$. Here we include 
the first nontrivial term and obtain the effective-Gaussian action 
$S_\mathrm{Gauss} = S_0 + S_2$, as the first-order term $S_1$ trivially vanishes 
thanks to the saddle-point condition given in the main text. 

For our interests, the zeroth-order action in $\Lambda_q$ can be written as
$
S_0 = \Delta_0^2/(T U) + \sum_{s \mathbf{k}} \xi_{s \mathbf{k}}/(2T)
- (1/2) \sum_{k} \ln [\det (\mathbf{G}_k^{-1}/T)],
$
where the collective index $k = (\mathbf{k}, \mathrm{i}\omega_\ell)$ denotes both the particle 
momentum $\mathbf{k}$ and the fermionic Matsubara frequency 
$\omega_\ell = (2\ell + 1)\pi T$~\cite{carlos97, iskin05, he12, shenoy12}. Here,
$
\mathbf{G}_k^{-1} = \mathrm{i}\omega_\ell \mathbf{1}  - H_\mathbf{k}^0
$
is the inverse Green's function for the saddle-point Hamiltonian $H_\mathbf{k}^0$, i.e.,
\begin{align*}
\mathbf{G}_k^{-1} = 
\left[
\begin{array}{cc}
 (\mathrm{i} \omega_\ell - \xi_{\mathbf{k}}) \sigma_0 
 - \mathbf{d}_\mathbf{k} \cdot \boldsymbol{\sigma} & \mathrm{i} \Delta_0 \sigma_y \\
  - \mathrm{i} \Delta_0 \sigma_y & (\mathrm{i}\omega_\ell + \xi_{\mathbf{k}}) \sigma_0 
  - \mathbf{d}_{\mathbf{k}} \cdot \boldsymbol{\sigma}^* \\
\end{array}
\right],
\end{align*}
determining the quasiparticle and quasihole energies through
$
\det \mathbf{G}_k^{-1} = [(\mathrm{i}\omega_\ell)^2 - E_{+,\mathbf{k}}^2]
[(\mathrm{i}\omega_\ell)^2 - E_{-,\mathbf{k}}^2] = 0,
$
leading to
$
E_{s \mathbf{k}} = \sqrt{\xi_{s\mathbf{k}}^2 + \Delta_0^2}
$
for the former. Performing the summation over the fermionic Matsubara frequency 
leads to the saddle-point action given in Eq.~(\ref{eqn:S0}) of the main text.

The second-order term in $\Lambda_q$ can be written as
$
S_2 = \sum_q |\Lambda_q|^2/(T U) + (1/4) \mathrm{Tr} \sum_{kq} \mathbf{G}_k \mathbf{\Sigma}_q 
\mathbf{G}_{k+q} \mathbf{\Sigma}_{-q},
$
where $\mathrm{Tr}$ denotes a trace over the particle/hole and spin sectors, 
and the collective index $q = (\mathbf{q}, \mathrm{i}\nu_n)$ denotes both the pair 
momentum $\mathbf{q}$ and the bosonic Matsubara frequency 
$\nu_n = 2\pi nT$~\cite{iskin05}. Here, the matrix elements of the Green's 
function $\mathbf{G}_k$ can be written as
\begin{align}
\label{eqn:G11}
G_k^{11} &= \frac{1}{2} \sum_s \frac{\mathrm{i} \omega_\ell + \xi_{s \mathbf{k}}}{(\mathrm{i} \omega_\ell)^2 - E_{s\mathbf{k}}^2} 
\left(\sigma_0 + s\widehat{\mathbf{d}}_\mathbf{k} \cdot \boldsymbol{\sigma} \right), \\
\label{eqn:G22}
G_k^{22}&= \frac{1}{2} \sum_s \frac{\mathrm{i} \omega_\ell - \xi_{s \mathbf{k}}}{(\mathrm{i} \omega_\ell)^2 - E_{s\mathbf{k}}^2} 
\left(\sigma_0 - s\widehat{\mathbf{d}}_\mathbf{k} \cdot \boldsymbol{\sigma}^* \right), \\
\label{eqn:G12}
G_k^{12} &= \frac{1}{2} \sum_s \frac{-\mathrm{i}\Delta_0\sigma_y}
{(\mathrm{i} \omega_\ell)^2 - E_{s\mathbf{k}}^2} 
\left(\sigma_0 - s\widehat{\mathbf{d}}_\mathbf{k} \cdot \boldsymbol{\sigma^*}\right), \\
\label{eqn:G21}
G_k^{21} &= \frac{1}{2} \sum_s \frac{\mathrm{i}\Delta_0\sigma_y}
{(\mathrm{i} \omega_\ell)^2 - E_{s\mathbf{k}}^2} 
\left(\sigma_0 + s\widehat{\mathbf{d}}_\mathbf{k} \cdot \boldsymbol{\sigma}\right),
\end{align}
and those of the fluctuations $\mathbf{\Sigma}_q$ are
$
\Sigma_q^{11} = \Sigma_q^{22} = 0,
$
$
\Sigma_q^{12} = \mathrm{i} \Lambda_q \sigma_y
$
and
$
\Sigma_q^{21} = -\mathrm{i} \Lambda_{-q}^* \sigma_y
$
~\cite{edelstein95, gorkov01}. After a lengthy but a straightforward summation over 
the fermionic Matsubara frequency, we find
$
S_2 = [1/(2 T)] \sum_q \mathbf{\bar{\Lambda}}_q^\dagger \mathbf{M}_q \mathbf{\bar{\Lambda}}_q,
$
where
$
\mathbf{\bar{\Lambda}}_q^\dagger = (\Lambda_q^* \, \Lambda_{-q})
$
and $\mathbf{M}_q$ is the inverse fluctuation 
propagator~\cite{carlos97, iskin05, he12, shenoy12}. Simplifying the notation
by denoting $\xi_{s, \mathbf{k}+\mathbf{q}/2}$ by $\xi$;
$\xi_{s', -\mathbf{k}+\mathbf{q}/2}$ by $\xi'$;
$E_{s, \mathbf{k}+\mathbf{q}/2}$ by $E$;
$E_{s', -\mathbf{k}+\mathbf{q}/2}$ by $E'$; and
$\widehat{\mathbf{d}}_{\pm \mathbf{k}+\mathbf{q}/2}$ by $\widehat{\mathbf{d}}_\pm$,
and defining the functions $u^2 = (1+\xi/E)/2$; $u'^2 = (1+\xi'/E')/2$; $v^2 = (1-\xi/E)/2$; 
$v'^2 = (1-\xi'/E')/2$; $f = 1/(e^{E/T}+1)$; and $f' = 1/(e^{E'/T}+1)$,
a compact way to write the matrix elements of $\mathbf{M}_q$ is~\cite{carlos97, iskin05}
\begin{align}
M_q^{11} &= \frac{1}{U} + \frac{1}{4} \sum_{ss' \mathbf{k}} 
\left(1 - ss' \widehat{\mathbf{d}}_+ \cdot \widehat{\mathbf{d}}_-\right) \nonumber \\
& \times \bigg[(1-f-f') \left( \frac{u^2 u'^2}{\mathrm{i}\nu_n-E-E'} - \frac{v^2 v'^2}{\mathrm{i}\nu_n+E+E'} \right) \nonumber \\
+ &(f-f') \left( \frac{v^2 u'^2}{\mathrm{i}\nu_n+E-E'} - \frac{u^2 v'^2}{\mathrm{i}\nu_n-E+E'} \right) 
\bigg], \\
M_q^{12} &= \frac{1}{4} \sum_{ss' \mathbf{k}} 
\left(1 - ss' \widehat{\mathbf{d}}_+ \cdot \widehat{\mathbf{d}}_-\right) \nonumber \\
& \times \bigg[(1-f-f') \left( \frac{u v u 'v'}{\mathrm{i}\nu_n+E+E'} - \frac{u v u'v'}{\mathrm{i}\nu_n-E-E'} \right) \nonumber \\
+ &(f-f') \left( \frac{u v u' v'}{\mathrm{i}\nu_n+E-E'} - \frac{u v u' v'}{\mathrm{i}\nu_n-E+E'} \right) 
\bigg].
\end{align}
The remaining elements are related via $M_q^{22} = M_{-q}^{11}$
and $M_q^{21} = M_{q}^{12}$. We note that while $M_q^{12}$ is even both 
under $\mathbf{q} \to -\mathbf{q}$ and $\mathrm{i}\nu_n \to -\mathrm{i}\nu_n$, $M_q^{11}$ is 
even only under $\mathbf{q} \to -\mathbf{q}$.

While calculating the collective excitations, it is useful to represent
$
\Lambda_q = \alpha_q e^{\mathrm{i} \gamma_q} 
$
in terms of real functions $\alpha_q$ and $\gamma_q$, and associate 
$
\lambda_q = \sqrt{2} \alpha_q \cos(\gamma_q)
$
with the amplitude degrees of freedom and
$
\theta_q = \sqrt{2}\alpha_q \sin(\gamma_q)
$
with the phase ones in the small $\gamma_q$ limit. This procedure 
corresponds to a unitary transformation
$
\Lambda_q = (\lambda_q + \mathrm{i} \theta_q)/\sqrt{2},
$
where $\lambda_q$ and $\theta_q$ are real functions. Furthermore, assuming 
$\lambda_{-q} = \lambda_q^*$ and $\theta_{-q} = \theta_q^*$, we find
\begin{align}
S_2 = \frac{1}{2T} \sum_q \mathrm{\Theta}_q^\dagger
\left( \begin{array}{cc}
M_{q,E}^{11}+M_q^{12} & \mathrm{i} M_{q,O}^{11} \\
- \mathrm{i} M_{q,O}^{11} & M_{q,E}^{11}-M_q^{12}
\end{array} \right)
\mathrm{\Theta}_q,
\end{align}
where we denote $\mathrm{\Theta}_q^\dagger = (\lambda_q^* \, \theta_q^*)$, 
and separate $M_q^{11} = M_{q,E}^{11} + M_{q,O}^{11}$ in terms of an even 
function
$
M_{q,E}^{11} = (M_q^{11} + M_q^{22})/2
$
in $\mathrm{i}\nu_n$ and an odd one
$
M_{q,O}^{11} = (M_q^{11} - M_q^{22})/2,
$
and use
$
M_q^{12} = M_q^{21}
$
~\cite{carlos97, iskin05, he12, shenoy12}.
The phase-only action given in Eq.~(\ref{eqn:S2theta}) is derived by integrating 
out the amplitude fields, and vice versa for the Eq.~(\ref{eqn:S2lambda}).
We note that since the quasiparticle-quasihole terms with the prefactor $(f-f')$ 
have the usual Landau singularity for $q \to (\mathbf{0}, 0)$ causing the collective 
modes to decay into the two-quasiparticle continuum, a small $q$ expansion 
is possible only in two cases: (i) just below the critical SF transition temperature 
provided $\Delta_0 \to 0 \ll \mathrm{i}\nu_n \to \omega$~\cite{iskin18b}, and (ii) at $T = 0$. 
In this work, we are interested in the latter case~\cite{earlierworks}.


\begin{thebibliography}{99}


\bibitem{provost80}
J. P. Provost and G. Vallee,
Riemannian structure on manifolds of quantum states,
Commun. Math. Phys. \textbf{76}, 289 (1980).

\bibitem{berry89}
M. V. Berry,
The quantum phase, five years after in Geometric Phases in Physics,
edited by A. Shapere and F. Wilczek (World Scientific, Singapore, 1989).

\bibitem{thouless98}
D. J. Thouless,
Topological Quantum Numbers in Nonrelativistic Physics,
(World   Scientific,  Singapore,1998)

\bibitem{xiao10}
D. Xiao, M.-C. Chang, and Q. Niu, 
Berry phase effects on electronic properties,
Rev. Mod. Phys. \textbf{82},1959 (2010).

\bibitem{qi11} X.-L. Qi and S.-C. Zhang, 
Topological insulators and superconductors,
Rev. Mod. Phys. \textbf{83}, 1057 (2011).


\bibitem{iskin18a}
M. Iskin,
Exposing the quantum geometry of spin-orbit coupled Fermi superfluids, 
Phys. Rev. A \textbf{97}, 063625 (2018).

\bibitem{iskin18b}
M. Iskin,
Quantum metric contribution to the pair mass in spin-orbit coupled Fermi superfluids,
Phys. Rev. A \textbf{97}, 033625 (2018). 


\bibitem{resta11}
R. Resta, 
The insulating state of matter: A geometrical theory,
Eur. Phys. J. B \textbf{79}, 121 (2011).



\bibitem{tan19}
X. Tan, D.-W. Zhang, Z. Yang, J. Chu, Y.-Q. Zhu, D. Li, X. Yang, S. Song, Z. Han, Z. Li, 
Y. Dong, H.-F. Yu, H. Yan, S.-L. Zhu, and Y. Yu,
Experimental Measurement of the Quantum Metric Tensor and Related Topological
Phase Transition with a Superconducting Qubit,
Phys. Rev. Lett. \textbf{122}, 210401 (2019).

\bibitem{yu18}
M. Yu, P. Yang, M. Gong, Q. Cao, Q. Lu, H. Liu, M. B. Plenio, F. Jelezko, T. Ozawa, 
N. Goldman, S. Zhang, and J. Cai,
Experimental measurement of the complete quantum geometry of a solid-state spin system,
arXiv:1811.12840 (2018).



\bibitem{peotta15}
S. Peotta and P. T\"{o}rm\"{a}, 
Superfluidity in topologically nontrivial flat bands,
Nat. Commun. {\bf 6}, 8944 (2015).

\bibitem{liang17}
L. Liang, T. I. Vanhala, S. Peotta, T. Siro, A. Harju, and P. T\"{o}rm\"{a},
Band geometry, Berry curvature, and superfluid weight,
Phys. Rev. B {\bf 95}, 024515 (2017).



\bibitem{hu19}
X. Hu, T. Hyart, D. I. Pikulin, and E. Rossi,
Geometric and conventional contribution to superfluid weight in twisted bilayer graphene,
arXiv:1906.07548 (2019).
 
\bibitem{julku19}
A. Julku, T. J. Peltonen, L. Liang, T. T. Heikkil\"a, and P. T\"orm\"a, 
Superfluid weight and Berezinskii-Kosterlitz-Thouless transition temperature 
of twisted bilayer graphene,
arXiv:1906.06313 (2019).

\bibitem{xie19}
F. Xie, Z. Song, B. Lian, and B. A. Bernevig,
Topology-Bounded Superfluid Weight In Twisted Bilayer Graphene,
arXiv:1906.02213 (2019).



\bibitem{wang12} P. Wang, Z. Yu, Z. Fu, J. Miao, L. Huang, S. Chai, H. Zhai, and J. Zhang, 
Spin-orbit coupled degenerate Fermi gases,
Phys. Rev. Lett. \textbf{109}, 095301 (2012).

\bibitem{cheuk12} L. W. Cheuk, A. T. Sommer, Z. Hadzibabic, T. Yefsah, W. S. Bakr, and M. W. Zwierlein, 
Spin-Injection Spectroscopy of a Spin-Orbit Coupled Fermi Gas,
Phys. Rev. Lett. \textbf{109}, 095302 (2012).
 
\bibitem{williams13} R. A. Williams, M. C. Beeler, L. J. LeBlanc, K. Jim\'enez-Garc\'ia, and I. B. Spielman, 
Raman-induced interactions in a single-component Fermi gas near an s-wave Feshbach resonance,
Phys. Rev. Lett. \textbf{111}, 095301 (2013).



\bibitem{huang16}
L. Huang, Z. Meng, P. Wang, P. Peng, S.-L. Zhang, L. Chen, D. Li, Q. Zhou, and J. Zhang,
Experimental realization of a two-dimensional synthetic spin-orbit coupling in ultracold Fermi gases,
Nat. Phys. \textbf{12}, 540 (2016).

\bibitem{meng16}
Z. Meng, L. Huang, P. Peng, D. Li, L. Chen, Y. Xu, C. Zhang, P. Wang, and J. Zhang,
Experimental Observation of a Topological Band Gap Opening in Ultracold Fermi Gases 
with Two-Dimensional Spin-Orbit Coupling,
Phys. Rev. Lett. \textbf{117}, 235304 (2016).

\bibitem{earlierworks} This work is complementary to our recent works. 
In~\cite{iskin18a}, we used a linear response theory and identified the SF-density 
tensor from the response of the thermodynamic potential to an infinitesimal 
superfluid flow. In~\cite{iskin18b}, we derived the time-dependent Ginzburg-Landau 
theory near the critical temperatures, and identified the effective-mass tensor 
of the Cooper pairs by making an analogy with the Gross-Pitaevskii theory of 
the corresponding molecular Bose gas. 



\bibitem{carlos97}
J. R. Engelbrecht, M. Randeria, and C. A. R. S\'a de Melo,
BCS to Bose crossover: Broken-symmetry state,
Phys. Rev. B \textbf{55}, 15153 (1997)

\bibitem{iskin05}
M. Iskin and C. A. R. S\'a de Melo,
BCS-BEC crossover of collective excitations in two-band superfluids,
Phys. Rev. B \textbf{72}, 024512 (2005).

\bibitem{he12}
L. He and X.-G. Huang,
BCS-BEC crossover in three-dimensional Fermi gases with spherical spin-orbit coupling,
Phys. Rev. B \textbf{86}, 014511 (2012).

\bibitem{shenoy12}
J. P. Vyasanakere and V. B. Shenoy,
Collective excitations, emergent Galilean invariance, and boson-boson interactions 
across the BCS-BEC crossover induced by a synthetic Rashba spin-orbit coupling,
Phys. Rev. A \textbf{86}, 053617 (2012).

%

\bibitem{collnote}
To be more precise, one needs to derive the characteristic equation up to 
fourth order in the expansion, e.g., 
$
\beta_5 \omega^4 + \sum_{ijkl} \beta_4^{ijkl} q_iq_jq_kq_l 
+ \sum_{ij} \beta_3^{ij} \omega^2 q_i q_j + \beta_2\omega^2 + \sum_{ij} \beta_1^{ij} q_iq_j = 0,
$
such that
$
x_{ij} = -\beta_1^{ij}/\beta_2,
$
$
\omega_0^2 = -\beta_2/\beta_5
$
and
$
y_{ij} = -\beta_3^{ij}/\beta_5 + \beta_1^{ij}/\beta_2.
$
While our quadratic expansion fully determines $x_{ij}$, we neglect the additional 
corrections to $\omega_0^2$ and $y_{ij}$ coming from the higher order terms.

\bibitem{sdMforms} The familiar expressions for the one-band coefficients 
can be written as~\cite{carlos97}
$
Q_{ij} = Q_{ij}^\mathrm{intra} = \sum_\mathbf{k} 
[ \xi_\mathbf{k} \partial^2 \xi_\mathbf{k}/(\partial k_i \partial k_j)
- (\partial \xi_\mathbf{k}/\partial k_i) (\partial \xi_\mathbf{k}/\partial k_j) 
(1- 3 \Delta_0^2 / E_\mathbf{k}^2) ]/(8 E_\mathbf{k}^3)
$
and 
$
C_{ij} = C_{ij}^\mathrm{intra} = \sum_\mathbf{k} 
[ \xi_\mathbf{k} (1 - 3\Delta_0^2/E_\mathbf{k}^2) \partial^2 \xi_\mathbf{k}/(\partial k_i \partial k_j)
- (\partial \xi_\mathbf{k}/\partial k_i) (\partial \xi_\mathbf{k}/\partial k_j) 
(1- 10 \Delta_0^2 \xi_\mathbf{k}^2 / E_\mathbf{k}^4) ]/(8 E_\mathbf{k}^3),
$
and our intraband coefficients recover them in the absence of a SOC 
when $d_\mathbf{k} \to 0$.

\bibitem{iskin19a}
M. Iskin, 
Superfluid stiffness for the attractive Hubbard model on a honeycomb optical lattice, 
Phys. Rev. A \textbf{99}, 023608 (2019).
 
 \bibitem{iskin19b}
M. Iskin,
Origin of flat-band superfluidity on the Mielke checkerboard lattice,
Phys. Rev. A \textbf{99}, 053608 (2019).

\bibitem{edelstein95}
V. M. Edelstein,
Magnetoelectric Effect in Polar Superconductors,
Phys. Rev. Lett. \textbf{75}, 2004 (1995).

\bibitem{gorkov01}
L. P. Gorkov and E. I. Rashba,
Superconducting 2D System with Lifted Spin Degeneracy: Mixed Singlet-Triplet State,
Phys. Rev. Lett. \textbf{87}, 037004 (2001).



\end{thebibliography}
\end{document}